\documentclass{article}
\usepackage[T1]{fontenc} 
\usepackage[utf8]{inputenc} 
\usepackage{ismir,amsmath,cite,url}
\usepackage{graphicx}
\usepackage{color}

\usepackage{soul} 
\usepackage{multirow}
\usepackage{amsfonts}

\usepackage{lineno}

\renewenvironment{description}[1][0pt]
  {\list{}{\labelwidth=.25cm \leftmargin=#1
   }}
  {\endlist}

\newcommand{\MASK}[1]{MASKED-FOR-REVIEW}

\usepackage{tikz}
\usetikzlibrary{fit,shapes.geometric}

\newcounter{nodemarkers}
\newcommand\circletext[1]{%
    \tikz[overlay,remember picture]
        \node (marker-\arabic{nodemarkers}-a) at (0,1.5ex) {};%
    #1%
    \tikz[overlay,remember picture]
        \node (marker-\arabic{nodemarkers}-b) at (0,0){};%
    \tikz[overlay,remember picture,inner sep=2pt]
        \node[draw,ellipse,fit=(marker-\arabic{nodemarkers}-a.center) (marker-\arabic{nodemarkers}-b.center)] {};%
    \stepcounter{nodemarkers}%
}

\newcommand{\DALI}{{DALI} }
\newcommand{\WASABI}{{WASABI} }

\title{Content based singing voice source separation via strong conditioning using aligned phonemes}

\twoauthors
   {Gabriel Meseguer-Brocal} {STMS UMR9912, Ircam/CNRS/SU, Paris \\ \tt\small  gabriel.meseguerbrocal@ircam.fr}
   {Geoffroy Peeters} {LTCI, Institut Polytechnique de Paris \\ \tt\small geoffroy.peeters@telecom-paris.fr}

\begin{document}

\maketitle
\begin{abstract}

Informed source separation has recently gained renewed interest with the introduction of neural networks and the availability of large multitrack datasets containing both the mixture and the separated sources.
These approaches use prior information about the target source to improve separation.
Historically, Music Information Retrieval researchers have focused primarily on score-informed source separation, but more recent approaches explore lyrics-informed source separation.
However, because of the lack of multitrack datasets with time-aligned lyrics, models use weak conditioning with non-aligned lyrics.
In this paper, we present a multimodal multitrack dataset with lyrics aligned in time at the word level with phonetic information as well as explore strong conditioning using the aligned phonemes.
Our model follows a {U-Net} architecture and takes as input both the magnitude spectrogram of a musical mixture and a matrix with aligned phonetic information.
The phoneme matrix is embedded to obtain the parameters that control Feature-wise Linear Modulation ({FiLM}) layers.
These layers condition the {U-Net} feature maps to adapt the separation process to the presence of different phonemes via affine transformations.
We show that phoneme conditioning can be successfully applied to improve singing voice source separation.

\end{abstract}



\section{Introduction}\label{sec:introduction}

Music source separation aims to isolate the different instruments that appear in an audio mixture (a mixed music track), reversing the mixing process.
Informed-source separation uses prior information about the target source to improve separation.
Researchers have shown that deep neural architectures can be effectively adapted to this paradigm~\cite{Kinoshita_2015, miron_2017}.
Music source separation is a particularly challenging task.
Instruments are usually correlated in time and frequency with many different harmonic instruments overlapping at several dynamics variations.
Without additional knowledge about the sources the separation is often infeasible.
To address this issue, Music Information Retrieval (MIR) researchers have integrated into the source separation process prior knowledge about the different instruments presented in a mixture, or musical scores that indicate where sounds appear.
This prior knowledge improves the performances~\cite{Slizovskaia_2020, Ewert_2014, miron_2017}. Recently, conditioning learning has shown that neural networks architectures can be effectively controlled for performing different music source isolation tasks~\cite{meseguerbrocal_2019, Tzinis_2019,Slizovskaia_2019, Seetharaman_2019, Samuel_2020, Schulze_2019}

Various multimodal context information can be used.
Although MIR researchers have historically focused on score-informed source separation to guide the separation process, lyrics-informed source separation has become an increasingly popular research area~\cite{Chandna_2020, Schulze_2019}.
Singing voice is one of the most important elements in a musical piece~\cite{demetriou_2018}.
Singing voice tasks (e.g. lyric or note transcription) are particularly challenging given its variety of timbre and expressive versatility.
Fortunately, recent data-driven machine learning techniques have boosted the quality and inspired many recent discoveries~\cite{gomez_2018, humphrey_2018}.
Singing voice works as a musical instrument and at the same time conveys a semantic meaning through the use of language~\cite{humphrey_2018}.
The relationship between sound and meaning is defined by a finite phonetic and semantic representations~\cite{goldsmith_1976, ladd_2008}.
Singing in popular music usually has a specific sound based on phonemes, which distinguishes it from the other musical instruments.
This motivates researchers to use prior knowledge such as a text transcript of the utterance or linguistic features to improve the singing voice source sparation~\cite{Chandna_2020, Schulze_2019}.
However, the lack of multitrack datasets with time-aligned lyrics has limited them to develop their ideas and only weak conditioning scenarios have been studied, i.e. using the context information without explicitly informing where it occurs in the signal.
Time-aligned lyrics provide abstract and high-level information about the phonetic characteristics of the singing signal.
This prior knowledge can facilitate the separation and be beneficial to the final isolation.


Looking for combining the power of data-driven models with the adaptability of informed approaches, we propose a multitrack dataset with time-aligned lyrics.
We explore then how we can use strong conditioning where the content information about the lyrics is available frame-wise to improve vocal sources separation.
We investigate strong and weak conditioning using the aligned phonemes via Feature-wise Linear Modulation ({FiLM}) layer~\cite{Perez_2017} in {U-Net} based architecture~\cite{Jansson_2017}.
We show that phoneme conditioning can be successfully applied to improve standard singing voice source separation and that simplest strong conditioning outperforms any other scenario.



\section{Related work}\label{sec:sota}

Informed source separation use context information about the sources to improve the separation quality, introducing in models additional flexibility to adapt to observed signals.
Researchers have explored different approaches for integrating different prior knowledge in the separation~\cite{liutkus_2013}.
Most of the recent data-driven music source separation methods use weak conditioning with prior knowledge about the different instruments presented in a mixture~\cite{Slizovskaia_2020, meseguerbrocal_2019, Slizovskaia_2019, Seetharaman_2019, Samuel_2020}.
Strong conditioning has been primarily used in score-informed source separation. In this section, we review works related to this topic as well as novel approaches that explore lyrics-informed source separation.

\subsection{Score-informed music source separation}

Scores provide prior knowledge for source separation in various ways. For each instrument (source), it defines which notes are played at which time, which can be linked to audio frames.
This information can be used to guide the estimation of the harmonics of the sound source at each frame~\cite{Ewert_2014, miron_2017}.
Pioneer approaches rely on non-negative matrix factorization ({NMF})~\cite{ewert_2012, duan_2011, rodriguez_2015, fritsch_2013}.
These methods assume that the audio is synchronized with the score and use different alignment techniques to achieve this.
%
Nevertheless, alignment methods introduce errors.
Local misalignments influence the quality of the separation~\cite{duan_2011, miron_2015}.
This is compensated by allowing a tolerance window around note onsets and offsets~\cite{ewert_2012, fritsch_2013} or with context-specific methods to refine the alignment~\cite{miron_2016}.
Current approaches use deep neural network architectures and filtering spectrograms by the scores and generating masks for each source~\cite{miron_2017}. The score-filtered spectrum is used as input to an encoder-decoder convolutional  neural  network (CNN) architecture similar to \cite{Chandna_2017}.
\cite{Ewert_2017} propose an unsupervised method where scores guide the representation learning to induce structure in the separation. They add class activity penalties and structured dropout extensions to the encoder-decoder architecture.
Class activity penalties capture the uncertainty about the target label value and structured dropout uses labels to enforce a specific structure, canceling activity related to unwanted note.

\subsection{Text-informed music source separation}

Due to the importance of singing voice in a musical piece~\cite{demetriou_2018}, it is one of the most useful source to separate in a music track.
Researchers have integrated the vocal activity information to constrain a robust principal component analysis (RPCA) method, applying a vocal/non-vocal mask or ideal time-frequency binary mask~\cite{Chan_2015}.
\cite{Schulze_2019}~propose a bidirectional recurrent neural networks (BRNN) method that includes context information extracted from the text via attention mechanism.
The method takes as input a whole audio track and its associated text information and learn alignment between mixture and context information that enhance the separation.
Recently, \cite{Chandna_2020} extract a representation of the linguistic content related to cognitively relevant features such as phonemes (but they do not explicitly predict the phonemes) in the mixture. The linguistic content guide the synthetization of the vocals.

\section{Formalization}\label{sec:formalization}
We use the multimodal information as context to guide and improve the separation.
We formalize our problem satisfying certain properties summarized as~\cite{Bengio_2013}:

\begin{description}[.15cm]
  \item[How is the multimodal model constructed?] We divide the model into two distinct parts~\cite{dumoulin_2018}: a \textit{generic} network that carries on the main computation and a \textit{control mechanism} that conditions the computation regarding context information and adds additional flexibility. The \textit{conditioning} itself is performed using {FiLM} layers~\cite{Perez_2017}. {FiLM} can effectively modulate a generic source separation model by some external information, controlling a single model to perform different instrument source separations~\cite{meseguerbrocal_2019, Slizovskaia_2020}.
  With this strategy, we can explore the \textit{control} and \textit{conditioning} parts regardless of the \textit{generic} network used.
  \item[Where is the context information used?] at which place in the \textit{generic} network we insert the context information, and defining how it affects the computation, i.e. weak (or strong) conditioning without (or with) explicitly informing where it occurs in the signal.
  \item[What context information?] We explore here prior information about the phonetic evolution of the singing voice, aligned in time with the audio. To this end, we introduce a novel multitrack dataset with lyrics aligned in time.
\end{description}


\section{Dataset}\label{sec:dataset}

The \DALI (Dataset of Aligned Lyric Information)~\cite{meseguerbrocal_2018} dataset is a collection of songs described as a sequence of time-aligned lyrics.
Time-aligned lyrics are described at four levels of granularity: \textbf{notes}, \textbf{words}, \textbf{lines} and \textbf{paragraphs}:


\begin{equation}\label{eq:dali-formal}
  A_{g} = (a_{k, g})_{k=1}^{K_g} \textrm{ where } a_{k,g} = (t_k^0, t_k^1, f_k, l_k, i_k)_g
\end{equation}

\noindent where $g$ is the granularity level and $K_g$ the number of elements of the aligned sequence, $t_k^0$ and $t_k^1$ being a text segment's start and end times (in seconds) with $t_k^0 < t_k^1$, $f_k$ a tuple $(f_\mathit{min}, f_\mathit{max})$ with the frequency range (in Hz) covered by all the notes in the segment (at the note level $f_\mathit{min} = f_\mathit{max}$, a vocal note), $l_k$ the actual lyric's information and $i_k = j$ the index that links an annotation $a_{k, g}$ with its corresponding upper granularity level annotation $a_{j, g+1}$.
The text segment's events for a song are ordered and non-overlapping - that is, $t_k^1 \le t_{k+1}^0 \forall k$.

There is a subset of \DALI of 513 multitracks with the \textit{mixture} and its separation in two sources, \textit{vocals} and \textit{accompaniment}. This subset comes from \WASABI dataset~\cite{meseguer_2017}.
The multitracks are distributed in 247 different artists and 32 different genres.
The dataset contains 35.4 hours with music and 14.1 hours with vocals, with a mean average duration per song of 220.83s and 98.97s with vocals.
All the songs are in English.

\begin{figure}[t]
  \centerline{
    \includegraphics[width=\columnwidth]{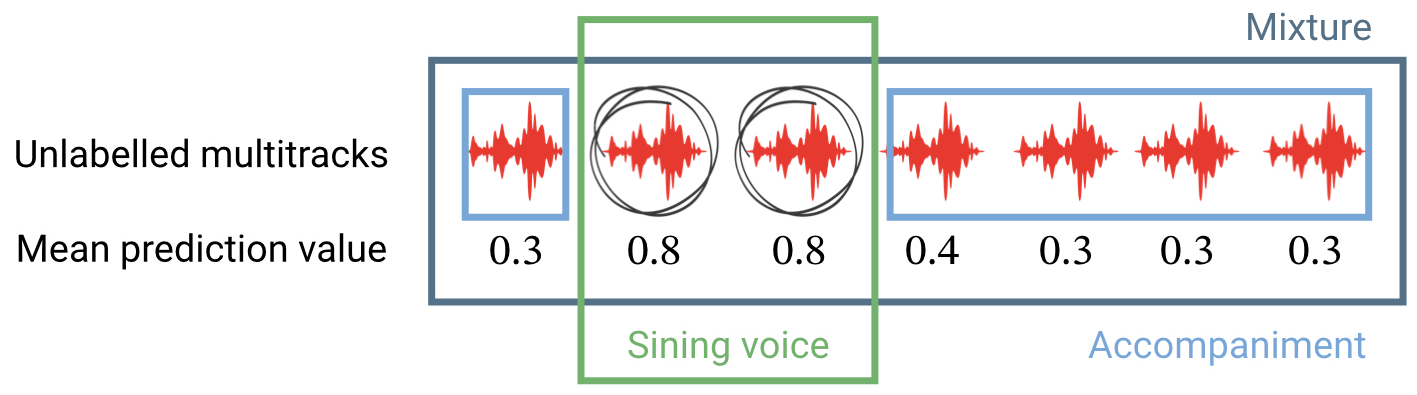}}
  \caption{Method used for creating the \textit{vocals}, \textit{accompaniment} and \textit{mixture} version.}
  \label{fig:multitracks}
\end{figure}

The original multitracks have the \textit{mixture} decomposed in a set of unlabeled sources in the form track\_1, track\_2, ..., track\_n. Depending of the songs, the files can be RAW (where each source is an instrument track e.g. a drum snare) or STEMS (where all the RAW files for an instrument are merged into a single file).
In the following, we explain how the \textit{vocals} and \textit{accompaniment} tracks are automatically created from these unlabelled sources. The process is summarized in~\figref{fig:multitracks}.

For each track $\tau$ of a multitrack song, we compute a singing voice probability vector overtime, using a pre-trained Singing Voice Detection (SVD) model~\cite{meseguerbrocal_2020}.
We obtain then a global mean prediction value per tracks $\epsilon_{\tau}$.
Assuming that there is at least one track with vocals,
we create the \textit{vocals} source by merging all the tracks with $\epsilon_{\tau} >=  \max_{\tau}(\epsilon_{\tau}) \cdot \nu$ where $\nu$ is a tolerance value set to $0.98$.
All the remaining tracks are fused to define the \textit{accompaniment}.
We manually checked the resulting sources.
The dataset is available at \url{https://zenodo.org/record/3970189}.

\begin{figure}[t]
  \centerline{
    \includegraphics[width=.9\columnwidth]{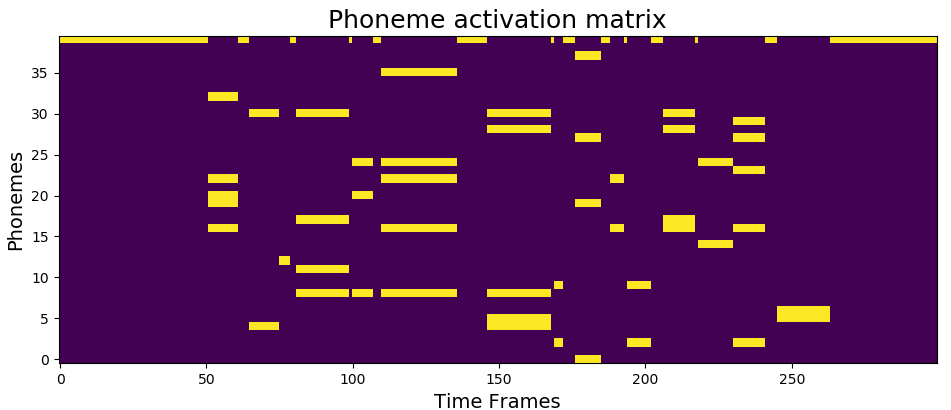}}
  \caption{Binary phoneme activation matrix. Note how words are represented as a bag of simultaneous phonemes.}
  \label{fig:phoneme}
\end{figure}

The second version of \DALI adds the phonetic information computed for the word level~\cite{meseguerbrocal_2020}.
This level has the words of the lyrics transcribed into a vocabulary of 39 different phoneme symbols as defined in the Carnegie Mellon Pronouncing Dictionary (CMUdict)\footnote{https://github.com/cmusphinx/cmudict}.
After selecting the desired time resolution, we can derive a time frame based phoneme context activation matrix $Z$, which is a binary matrix that indicates the phoneme activation  over time.
We add an extra row with the 'non-phoneme' activation with $1$ at time frames with no phoneme activation and $0$ otherwise. \figref{fig:phoneme} illustrates the final activation matrix.

%
Although we work only with phonemes per word information, we can derive similar activation matrices for other context information such as notes or characters.


\begin{table*}
\centering
\footnotesize
\begin{tabular}{r|c|c|c|c|c|c|c|c|c|c|c}
Model      & U-Net & $\mathit{W}_\mathit{si}$ & $\mathit{W}_\mathit{co}$ & $\mathit{S}_\mathit{a}$ & $\mathit{S}_\mathit{a*}$ & $\mathit{S}_\mathit{c}$ & $\mathit{S}_\mathit{c*}$ & $\mathit{S}_\mathit{f}$ & $\mathit{S}_\mathit{f*}$  & $\mathit{S}_\mathit{s}$ & $\mathit{S}_\mathit{s*}$    \\
\hline
$\theta$ &    $9.83 \cdot 10^6$     &   $+14,060$     &  $+2.35 \cdot 10^6$    &  $+1.97 \cdot 10^6$ & $+327,680$     &   $+80,640$   & $+40,960$   &  $+40,320$  & $+640$   & $+480$ & $+80$
\end{tabular}
\caption{Number of parameters ($\theta$) for the different configurations. We indicate increment to the U-Net architecture.}
\label{table:param}
\end{table*}

\section{Methodology}\label{sec:vunet_method}

Our method adapts the {C-U-Net} architecture~\cite{meseguerbrocal_2019} to the singing voice separation task, exploring how to use the prior knowledge defined by the phonemes to improve the vocals separation.

\textbf{Input representations}. Let $X \in \mathbb{R}^{T \times M}$  be the magnitude of the Short-Time Fourier Transform (STFT) with $M=512$ frequency bands and $T$ time frames.
We compute the STFT on an audio signal down-sampled at 8192~Hz using a window size of 1024 samples and a hop size of 768 samples.
Let $Z \in \mathbb{R}^{T \times P}$ be the aligned phoneme activation matrix with $P=40$ phoneme types and $T$ the same time frames as in $X$.
Our model takes as inputs two submatrix $x \in \mathbb{R}^{N \times M}$ and $z \in \mathbb{R}^{N \times P}$ of $N=128$ frames (11 seconds) derived from $X$ and $Z$.

\textbf{Model}. The {C-U-Net} model has two components (see \cite{meseguerbrocal_2019} for a general overview of the architecture): a \textbf{conditioned network} that processes $x$ and a \textbf{control mechanism} that conditions the computation with respect to $z$.
We denote by $x_d \in \mathbb{R}^{W \times H \times C}$ the intermediate features of the \textbf{conditioned network}, at a particular depth $d$ in the architecture.
$W$ and $H$ represent the `time' and `frequency' dimension and $C$ the number of feature channels (or feature maps).
A \textit{{FiLM}} layer conditions the network computation by applying an affine transformation to $x_d$:

\begin{equation}
    \label{eq_film}
    \mathit{FiLM}(x_d) = \gamma_{d}(z) \odot x_d + \beta_{d}(z)
\end{equation}

\noindent where $\odot$ denotes the element-wise multiplication and $\gamma_{d}(z)$ and $\beta_{d}(z)$ are learnable parameters with respect to the input context $z$.
A \textit{{FiLM}} layer can be inserted at any depth of the original model and its output has the same dimension as the  $x_d$ input, i.e. $\in \mathbb{R}^{W \times H \times C}$.
To perform \eqnref{eq_film}, $\gamma_{d}(z)$ and $\beta_{d}(z)$ must have the same dimensionaly as $x_d$, i.e. $\in \mathbb{R}^{W \times H \times C}$.
However, we can define them omitting some dimensions. This results in a non-matching dimensionality with $x_d$, solved by broadcasting (repeating) the existing information to the missing dimensions.

As in~\cite{Jansson_2017, Stoller_2018, meseguerbrocal_2019}, we use the U-Net\cite{Jansson_2017} as \textbf{conditioned network}, which has an encoder-decoder mirror architecture based on CNN blocks with skip connections between layers at the same hierarchical level in the encoder and decoder.
Each convolutional block in the encoder halves the size of the input and doubles the number of channels.
The decoder is made of a stack of transposed convolutional operation, its output has the same size as the input of the encoder.
Following the original {C-U-Net} architecture, we insert the \textit{{FiLM}} layers at each encoding block after the batch normalization and before the Leaky ReLU~\cite{meseguerbrocal_2019}.

We explore now the different \textbf{control mechanism} we use for conditioning the U-Net.

\subsection{Control mechanism for weak conditioning}

Weak conditioning refers to the cases where
\begin{itemize}
  \item $\gamma_{d}(z)$ and $\beta_{d}(z)$  $\in \mathbb{R}^1$: they are scalar parameters applied independently of the times $W$, the frequencies $H$ and the channel $C$ dimensions. They depend only on the depth $d$ of the layer within the network~\cite{meseguerbrocal_2019}.
  \item  $\gamma_{d}(z)$ and $\beta_{d}(z)$  $\in \mathbb{R}^C$: this is the original configuration proposed by~\cite{Perez_2017} with different parameters for each channel $c \in {1, ..., C}$.
\end{itemize}

We call them \textit{{FiLM} simple} ($\mathit{W}_\mathit{si}$) and \textit{{FiLM} complex} ($\mathit{W}_\mathit{co}$) respectively. Note how they apply the same transformation without explicitly informing where it occurs in the signal (same value over the dimension $W$ and $H$).




Starting from the context matrix $z \in \mathbb{R}^{N \times P}$, we define the \textbf{control mechanism} by first apply  the autopool layer proposed by~\cite{mcfee_2018}\footnote{The auto-pool layer is a tuned soft-max pooling that automatically adapts the pooling behavior to interpolate between mean and max-pooling for each dimension} to reduce the input matrix to a time-less vector.
We then fed this vector into a dense layer and two dense blocks each composed by a dense layer, 50\% dropout and batch normalization.
For \textit{{FiLM} simple}, the number of units of the dense layers are 32, 64 and 128.
For \textit{{FiLM} simple}, they are 64, 256 and 1024.
All neurons have ReLU activations.
The output of the last block is then used to feed two parallel and independent dense layer with linear activation which outputs all the needed $\gamma_{d}(z)$ and $\beta_{d}(z)$.
While for the \textit{{FiLM} simple} configuration we only need $12$ $\gamma_{d}$ and $\beta_{d}$ (one $\gamma_{d}$ and $\beta_{d}$ for each of the 6 different encoding blocks) for the \textit{{FiLM} complex} we need $2016$ (the encoding blocks feature channel dimensions are $16$, $32$, $64$, $128$, $256$ and $512$, which adds up to $1008$).

\subsection{Control mechanism for strong conditioning}

In this section, we extend the original \textit{{FiLM}} layer mechanism to adapt it to the strong conditioning scenario.

The context information represented in the input matrix $z$ describes the presence of the phonemes $p \in \{1,\ldots,P\}$ over time $n \in \{1, \ldots N\}$.
As in the popular Non-Negative Matrix factorization~\cite{lee_2001} (but without the non-negativity constraint), our idea is to represent this information as the product of tensors: an activation and two basis tensors.

The \textbf{activation tensor $z_d$} indicates which phoneme occurs at which time: $z_{d} \in \mathbb{R}^{W \times P}$  where $W$ is the dimension which represents the time at the current layer $d$ (we therefore need to map the time range of $z$ to the one of the layer $d$) and $P$ the number of phonemes.

\begin{figure}[t]
  \centerline{
    \includegraphics[width=.9\columnwidth]{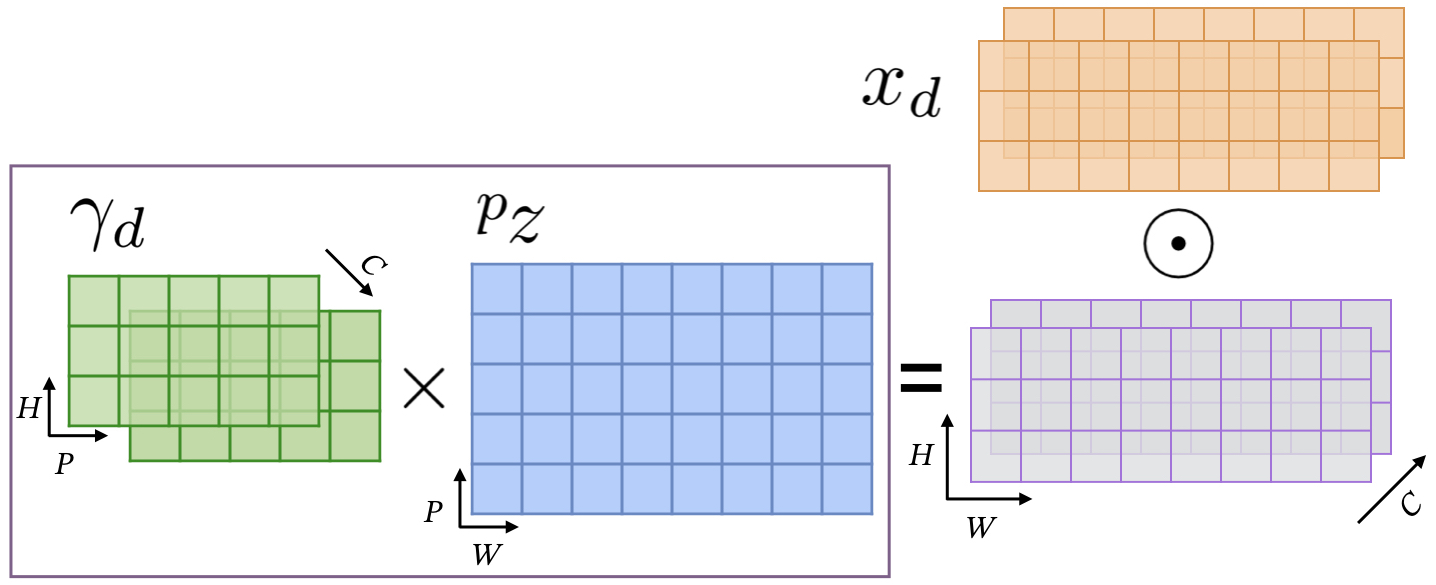}}
  \caption{Strong conditioning example with $(\gamma_{d} \times{} z_d)\odot x_d$. The phoneme activation $z_d$ defines how the basis tensors ($\gamma_d$) are employed for performing the conditioning on $x_d$.}
  \label{fig:strong}
\end{figure}

The \textbf{two basis tensors $\gamma_d$ and $\beta_d$} $\in \mathbb{R}^{H \times C \times P}$  where $H$ is the dimension which represents the frequencies at the current layer $d$, $C$ the number of input channels and $P$ the number of phonemes.
In other words, each phoneme $p$ is represented by a matrix in $\mathbb{R}^{H \times C}$ derived from \eqnref{eq:dali-formal}.
This matrix represents the specific conditioning to apply to $x_d$ if the phoneme exists (see~\figref{fig:strong}).
These matrices are learnable parameters (neurons with linear activations) but they do not depend on any particular input information (at a depth $d$ they do not depend on $x$ nor $z$), they are rather ``activated'' by $z_d$ at specific times.
As for the `weak`conditionning, we can define different versions of the tensors
\begin{itemize}
  \item the \textbf{all-version} ($\mathit{S}_\mathit{a}$) described so far with three dimensions: $\gamma_d , \beta_d \in \mathbb{R}^{H \times C \times P}$
  \item the \textbf{channel-version} ($\mathit{S}_\mathit{c}$): each phoneme is represented by a vector over input channels (therefore constant over frequencies): $\gamma_d , \beta_d \in \mathbb{R}^{C \times P}$
  \item the \textbf{frequency-version} ($\mathit{S}_\mathit{f}$): each phoneme is represented by a vector over input frequencies (therefore constant over channels): $\gamma_d , \beta_d \in \mathbb{R}^{H \times P}$
  \item the \textbf{scalar-version} ($\mathit{S}_\mathit{s}$): each phoneme is represented as a scalar (therefore constant over frequencies and channels): $\gamma_d , \beta_d \in \mathbb{R}^{P}$
\end{itemize}

\noindent The global conditioning mechanism can then be written as
\begin{equation} \label{eq:act_mul}
  \mathit{FiLM}(x_d, z_d) = (\gamma_{d} \times{} z_d) \odot x_d + (\beta_{d} \times{} z_d)
\end{equation}

\noindent where $\odot$ is the element-wise multiplication and $\times$ the matrix multiplication.
We broadcast $\gamma_d$ and $\beta_d$ for missing dimensions and transpose them properly to perform the matrix multiplication.
We test two different configurations: inserting {FiLM} at each encoder block as suggested in~\cite{meseguerbrocal_2019} and inserting {FiLM} only at the last encoder block as proposed at~\cite{Slizovskaia_2020}. We call the former `complete' and the latter `bottleneck' (denoted with $*$ after the model acronym).
We resume the different configurations at~\tabref{table:param}.

\section{Experiments}\label{sec:exp}


\begin{table}
\centering
\small
\begin{tabular}{c|c|c|c}
          & Train & Val & Test  \\
\hline
Threshold   &  $ .88  > \eta >= .7$  &  $.89 > \eta >= .88$     &   $.89  > \eta $      \\
\hline
Songs       & 357   &   30   & 101
\end{tabular}
\caption{\DALI split according to agreement score $\eta$.}
\label{table:split}
\end{table}

\textbf{DATA.} We split \DALI into three sets according to the normalized agreement score $\eta$ presented in~\cite{meseguerbrocal_2018} (see~\tabref{table:split}).
This score provides a global indication of the global alignment correlation between the annotations and the vocal activity.
\\~\\
\noindent \textbf{DETAILS.}
We train the model using batches of $128$ spectrograms randomly drawn from the training set with $1024$ batches per epoch.
The loss function is the mean absolute error between the predicted vocals (masked input mixture) and the original vocals. We use a learning rate of $0.001$ and the reduction on plateau and early stopping callbacks evaluated on the validation set, using patience of $15$ or $30$ respectively and a min delta variation for early stopping to $1e-5$.
Our output is a Time/Frequency mask to be applied to the magnitude of the input STFT mixture. We use the phase of the input STFT mixture to reconstruct the waveform with the inverse STFT algorithm.

For the strong conditioning, we apply a {\texttt{softmax}} on the input phoneme matrix $z$ over the phoneme dimension $P$ to constrain the outputs to sum to $1$, meaning it lies on a simplex, which helps in the optimization.

\begin{table}
\centering
\begin{tabular}{c|c|c|c|c|c}
Training & Test  & Aug & SDR & SIR & SAR \\

\hline\hline
Musdb18     & Musdb18   & False        & 4.27        & 13.17    & 5.17    \\
(90)        & (50)      & True         & 4.46        & 12.62    & 5.29    \\
\hline\hline
\multirow{4}{*}{\begin{tabular}[c]{@{}l@{}} DALI \\ (357) \end{tabular}}     & Musdb18     & False        & 4.60        & 14.03    & 5.39    \\
                          &           (50)                  & True         & 4.96        & 13.50    & 5.92    \\
\cline{2-6}
                          & DALI       & False        & 3.98        & 12.05    & 4.91    \\
                          & (101)      & True         & 4.05        & 11.40    & 5.32    \\
\end{tabular}
\caption{Data augmentation experiment.}
\label{table:data_aug}
\end{table}

\subsection{Evaluation metrics}
We evaluate the performances of the separation using the mir evaltoolbox~\cite{Raffel_2014}.
We compute three metrics: Source-to-Interference Ratios (SIR), Source-to-Artifact Ratios (SAR), and Source-to-Distortion Ratios (SDR)~\cite{Vincent_2006}.
In practice, SIR measures the interference from other sources, SAR the algorithmic artifacts introduce in the process and SDR resumes the overall performance.
We obtain them globally for the whole track.
However, these metrics are ill-defined for silent sources and targets.
Hence, we compute also the Predicted Energy at Silence (PES) and Energy at Predicted Silence (EPS) scores~\cite{Schulze_2019}. PES is the mean of the energy in the predictions at those frames with silent target and EPS is the opposite, the mean of the target energy of all frames with silent prediction and non-silent target.
For numerical stability, in our implementation, we add a small constant $\epsilon = 10^{-9}$ which results in a lower boundary of the metrics to be $-80$ dB~\cite{Slizovskaia_2020}.
We consider as silent segments those that have a total sum of less than $-25$ dB of the maximum absolute in the audio.
We report the \textbf{median} values of these metrics over the all tracks in the \DALI test set.
For SIR, SAR, and SDR larger values indicate better performance, for PES and EPS smaller values, mean better performance.

\subsection{Data augmentation}
Similarly as proposed in~\cite{uhlich_2017}, we randomly created `fake' input mixtures every $4$ real mixtures.
In non-augmented training, we employ the mixture as input and the vocals as a target.
However, this does not make use of the accompaniment (which is only employed during evaluation).
We can integrate it creating `fake' inputs by automatically mixing (mixing meaning simply adding) the target vocals to a random sample accompaniment from our training set.

We test the data augmentation process using the standard {U-Net} architecture to see whether it improves the performance (see~\tabref{table:data_aug}).
We train two models on \DALI and Musdb18 dataset~\cite{musdb_2018}\footnote{We use 10 songs of the training set for the early stopping and reduction on plateau callbacks}.
This data augmentation enables models to achieve better SDR and SAR but lower SIR.
Our best results ($4.96$ db SDR) are not state-of-the-art where the best-performing models on Musdb18 achieve (approximately $6.60$ db SDR)~\cite{stoter_2018}.

This technique does not reflect a large improvement when the model trained on \DALI is tested on \DALI. However, when this model is tested on Musdb18, it shows a better generalization (we have not seen any song of Musidb18 during training) than the model without data augmentation (we gain $0.36$ dB).
One possible explanation for not having a large improvement on \DALI testset is the larger size of the test set. It also can be due to the fact that vocal targets in \DALI still contain leaks such as low volume music accompaniment that come from the singer headphones.
We adopt this technique for training all the following models.

Finally, we confirmed a common belief that training with a large dataset and clean separated sources improves the separation over a small dataset~\cite{Pretet_2019}.
Both models trained on \DALI (with and without augmentation) improve the results obtained with the models trained on Musdb18.

Since we cannot test the conditioning versions on Musdb (no aligned lyrics), the results on the \DALI test ($4.05$ dB SDR) serves as a baseline to measure the contribution of the conditioning techniques (our main interest).

\begin{figure*}[t]
  \centerline{
    \includegraphics[width=\textwidth]{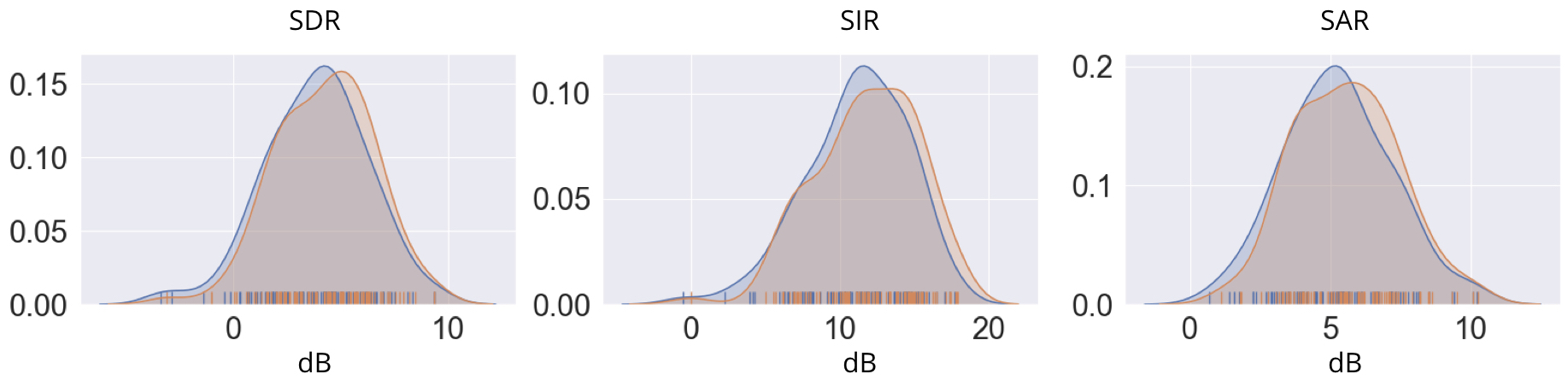}}
  \caption{Distribution of scores for the the standar {U-Net} (Blue) and $\mathit{S}_\mathit{s}$ (Orange).}
  \label{fig:comparing}
\end{figure*}

\begin{table}
\centering
\begin{tabular}{c|c|c|c|c|c}
Model & SDR & SIR & SAR & PES & EPS  \\
\hline\hline
{U-Net}                     & 4.05  & 11.40    &  5.32   &  -42.44   &   -64.84   \\
\hline
\hline
$\mathit{W}_\mathit{si}$  & \textbf{4.24}  & \textbf{11.78}    & 5.38   &  \textbf{-49.44}    &  -65.47    \\
\hline
$\mathit{W}_\mathit{co}$  & \textbf{4.24}  & \textbf{12.72}    & 5.15   & \textbf{-59.53}    & -63.46     \\
\hline
\hline
$\mathit{S}_\mathit{a}$  & 4.04    & \textbf{12.14}  & 5.13   & \circletext{\textbf{-59.68}}   & -61.73    \\
$\mathit{S}_\mathit{a*}$ & \textbf{4.27}    & \textbf{12.42}   & 5.26    & \textbf{-54.16}   & -64.56     \\
\hline
$\mathit{S}_\mathit{c}$  & \textbf{4.36}    & \textbf{12.47}   & 5.34    & \textbf{-57.11}   & -65.48   \\
$\mathit{S}_\mathit{c*}$ & \textbf{4.32}  & \textbf{12.86}    &  5.15    &  \textbf{-54.27}   &  \textbf{-66.35}    \\
\hline
$\mathit{S}_\mathit{f}$  & 4.10    & 11.40   & 5.24    & \-47.75   & -62.76    \\
$\mathit{S}_\mathit{f*}$ & 4.21       & \circletext{\textbf{13.13}}    & 5.05    & \textbf{-48.75}    &  \circletext{\textbf{-72.40}}     \\
\hline
$\mathit{S}_\mathit{s}$  & \textbf{\circletext{4.45}}    & \textbf{11.87}  &  \circletext{\textbf{5.52}}   & \textbf{-51.76}   & -63.44 \\
$\mathit{S}_\mathit{s*}$ & \textbf{4.26}      & \textbf{12.80}    &  5.25   & \textbf{-57.37}    &   -65.62  \\
\end{tabular}
\caption{Median performance in dB of the different models on the \DALI test set.
In bold are the results that significantly improve over the {U-Net} (p < 0.001) and inside the circles the best results for each metric.
}
\label{table:results}
\end{table}

\section{Results}\label{sec:results}

We report the median source separation metrics (SDR, SAR, SIR, PES, ESP) in~\tabref{table:results}. To measure the significance of the improvement differences, we performed a paired t-test between each conditioning model and the standard {U-Net} architecture, the baseline. This test measures ($p$-value) if the differences could have happened by chance. A low $p$-value indicates that data did not occur by chance.
As expected, there is a marginal (but statistical significance) improvement over most of the proposed methods, with a generalized $p<0.001$ for the SDR, SIR, and PES, except for the versions where the basis tensors have a `frequency' $H$ dimension.
This is an expected result since when singing, the same phoneme can be sung at different frequencies (appearing at many frequency positions in the feature maps). Hence, these versions have difficulties to find generic basis tensors.
This also explains why the `bottleneck' versions (for both $\mathit{S}_\mathit{f*}$ and $\mathit{S}_\mathit{a*}$) outperforms the `complete' while this is not the case for the other versions.
Most versions also improve the performance on silent vocal frames with a much lower PES.
However, there is no difference in predicting silence at the right time (same EPS).
The only metric that does not consistently improve is SAR, which measures the algorithmic artifacts introduced in the process.
Our conditioning mechanisms can not reduce the artifacts that seem more dependent on the quality of the training examples (it is the metric with higher improvement in the data augmentation experiment~\tabref{table:data_aug}).
\figref{fig:comparing} shows a comparison with the distribution of SDR, SIR, and SAR for the best model $\mathit{S}_\mathit{s}$ and the {U-Net}. We can see how the distributions move toward higher values.

One relevant remark is the fact that we can effectively control the network with just a few parameters.
$\mathit{S}_\mathit{s}$ just adds $480$ (or just $80$ for $\mathit{S}_\mathit{s*}$) new learnable parameters and have significantly better performance than $\mathit{S}_\mathit{a}$ that adds $1.97 \cdot 10^6$.
We believe that the more complex control mechanisms tend to find complex basis tensors that do not generalize well. In our case, it is more effective to perform a simple global transformation.
In the case of weak conditioning, both models behave similarly although $\mathit{W}_\mathit{si}$ has $1.955 \cdot 10^6$ fewer parameters than $\mathit{W}_\mathit{co}$.
This seems to indicate that controlling channels is not particularly relevant.

Regarding the different types of conditioning, when repeating the paired t-test between weak and strong models only $\mathit{S}_\mathit{s}$ outperforms the weak systems.
We believe that strong conditioning can lead to higher improvements but several issues need to be addressed. First, there are misalignments in the annotations that force the system to perform unnecessary operations which damages the computation. This is one of the possible explanations of why models with fewer parameters perform better. They are forced to find more generic conditions. The weak conditioning models are robust to these problems since they process $z$ and compute an optimal modification for a whole input patch (11s). We also need to
``disambiguate'' the phonemes inside words since they occur as a bag of phonemes at the same time (no individual onsets per phonemes inside one word, see \figref{fig:phoneme}). This prevents strong conditioning models to properly learn the phonemes in isolation, instead, they consider them jointly with the other phonemes.



\section{Conclusions}\label{sec:conclusions}
The goal of this paper is twofold. First, to introduce a new multimodal multitrack dataset with lyrics aligned in time.
Second, to improve singing voice separation using the prior knowledge defined by the phonetic characteristics.
We use the phoneme activation as side information and show that it helps in the separation.

In future works, we intend to use other prior aligned knowledge such as vocal notes or characters also defined in \DALI.
Regarding the conditioning approach and since it is transparent to the conditioned network, we are determined to explore recent state-of-the-art source separation methods such as {Conv-Tasnet}~\cite{Luo_2019}.
The current formalization of the two basis tensors $\gamma_d$ and $\beta_d$ does not depend on any external factor.
A way to exploit a more complex control mechanisms is to make these basis tensors dependent on the input mixture $x$ which may add additional flexibility.
Finally, we plan to jointly learn how to infer the alignment and perform the separation~\cite{Schulze_2020, Takahashi_2020}.

The general idea of lyrics-informed source separation leaves room for many possible extensions.
The present formalization relies on time-aligned lyrics which is not the real-world scenario.
Features similar to the phoneme activation~\cite{vaglio_2020, stoller_2019} can replace them or be used to align the lyrics as a pre-processing step. This two options adapts the current system to the real-world scenario.
These features can also help in properly placing and disambiguating the phonemes of a word to improve the current annotations.

\pagebreak

\textbf{Acknowledgement.}
This research has received funding from the French National Research Agency under the contract ANR-16-CE23-0017-01 (WASABI project). Implementation available at https://github.com/gabolsgabs/vunet

\bibliography{main}

\end{document}